\documentclass[letterpaper]{article} 
\usepackage[preprint]{aaai2027}  
\usepackage[hyphens]{url}  
\usepackage{graphicx} 
\usepackage{natbib}  
\usepackage{caption} 
\usepackage{algorithm}
\usepackage{algorithmic}
\usepackage{amsmath}
\usepackage{amssymb}
\usepackage{booktabs}
\usepackage{multirow}
\usepackage{graphicx}
\usepackage{amsthm}
\usepackage{subcaption}
\usepackage{newfloat}
\usepackage{listings}
\DeclareCaptionStyle{ruled}{labelfont=normalfont,labelsep=colon,strut=off} 
\floatstyle{ruled}
\newfloat{listing}{tb}{lst}{}
\floatname{listing}{Listing}

\newtheorem{assumption}{Assumption}
\newtheorem{theorem}{Theorem}
\newtheorem{proposition}{Proposition}
\newtheorem{corollary}{Corollary}
\newtheorem{remark}{Remark}

\usepackage{booktabs}

\title{CDBG: Causally Motivated Dual-Invariance Learning against Topological and Predictive Shifts in EEG Workload Recognition}
\author{
Yuzhe Zhang\textsuperscript{\rm 1}\corresponding,
Wenmin Zhou\textsuperscript{\rm 1},
Chengxi Xie\textsuperscript{\rm 2},
Kai He\textsuperscript{\rm 3},
Jihong Wang\textsuperscript{\rm 4}\corresponding,
Huan Liu\textsuperscript{\rm 4},
Man Yao\textsuperscript{\rm 5},
Daoqiang Zhang\textsuperscript{\rm 1}
}

\affiliations{
\textsuperscript{\rm 1}College of Artificial Intelligence,
Nanjing University of Aeronautics and Astronautics\\
\textsuperscript{\rm 2}School of Intelligent Science and Engineering,
Harbin Institute of Technology (Shenzhen)\\
\textsuperscript{\rm 3}School of Public Health,
National University of Singapore\\
\textsuperscript{\rm 4}School of Computer Science and Technology,
Xi'an Jiaotong University\\
\textsuperscript{\rm 5}Institute of Automation,
Chinese Academy of Sciences\\

}

\begin{document}

\maketitle

\begin{abstract}
Generalizing Electroencephalography (EEG)-based mental workload recognition to unseen subjects remains a formidable challenge due to severe inter-subject variability. While functional brain graphs effectively model distributed cognitive dynamics, their inherent subject-specificity induces two coupled distribution shifts: a \emph{class-conditional topological shift} in the underlying functional connectivity, and a \emph{predictive mechanism shift} in the learned representation-to-label mapping. Motivated by the subject-induced distribution shifts, we propose \textbf{CDBG}, a \textbf{C}ausally motivated \textbf{D}ual-invariance learning framework for \textbf{B}rain \textbf{G}raphs. CDBG disentangles and mitigates these shifts via a two-stage rationale learning pipeline. First, it employs stochastic edge masking to extract sparse, workload-predictive graph rationales, regularized by workload-conditional Laplacian spectral alignment to enforce topological invariance across subjects. Second, it applies subject-wise Invariant Risk Minimization (IRM) to the graph representations, ensuring environment-wise risk stationarity. Extensive experiments on a self-built air traffic controller EEG cognitive workload dataset and multiple public datasets under a strict leave-one-subject-out protocol demonstrate that CDBG significantly outperforms state-of-the-art cross-subject and graph-based baselines, improving the Macro-F1 score by up to 4.23\%, while simultaneously providing neurophysiologically interpretable functional rationales.
\end{abstract}


\section{Introduction}

Reliable mental workload recognition is a cornerstone for adaptive human-machine systems~\cite{charles2019mental}. While Electroencephalography (EEG) provides a non-invasive window into cognitive dynamics, decoding these signals is challenging due to their non-stationary nature, low signal-to-noise ratio, and large inter-subject variability~\cite{lotte2018review,wang2024eegpt}. Crucially, cognitive responses emerge from distributed coordination across cortical regions rather than isolated electrodes~\cite{bastos2016connectivity}. Functional brain graphs capture these interactions by modeling channel-wise activity and connectivity, and their topological organization inherently varies with task demand~\cite{dimitriadis2015workload,zhang2017graphworkload}. Consequently, graph representation learning has emerged as a powerful paradigm for EEG workload recognition~\cite{klepl2024survey,safari2024classification}.

Despite these advances, real-world deployment is severely hindered by the cross-subject generalization problem. Individual physiological differences produce significant distribution shifts, degrading performance on unseen subjects~\cite{albuquerque2022shifts}. Optimizing empirical risk over pooled source cohorts often encourages models to exploit subject-specific connectivity rather than task-intrinsic neural signatures. From a causal perspective, the observed functional brain graph $G$ is a downstream response jointly modulated by mental workload $Y$ and subject-related factors $S$, summarized as $Y\rightarrow G\leftarrow S$. Workload recognition therefore constitutes an anti-causal prediction problem: inferring $Y$ from its subject-modulated neural response $G$. Since $G$ conflates task-related responses and subject-dependent variations, predictive associations observed in source subjects may not generalize. Thus, we formulate this as causally motivated dual-invariance learning, seeking a sparse, workload-predictive subgraph $\widetilde G$ whose topology and induced predictive relationship remain stable across subject environments.

\begin{figure}[!t]
\centering
\includegraphics[width=0.9\columnwidth]{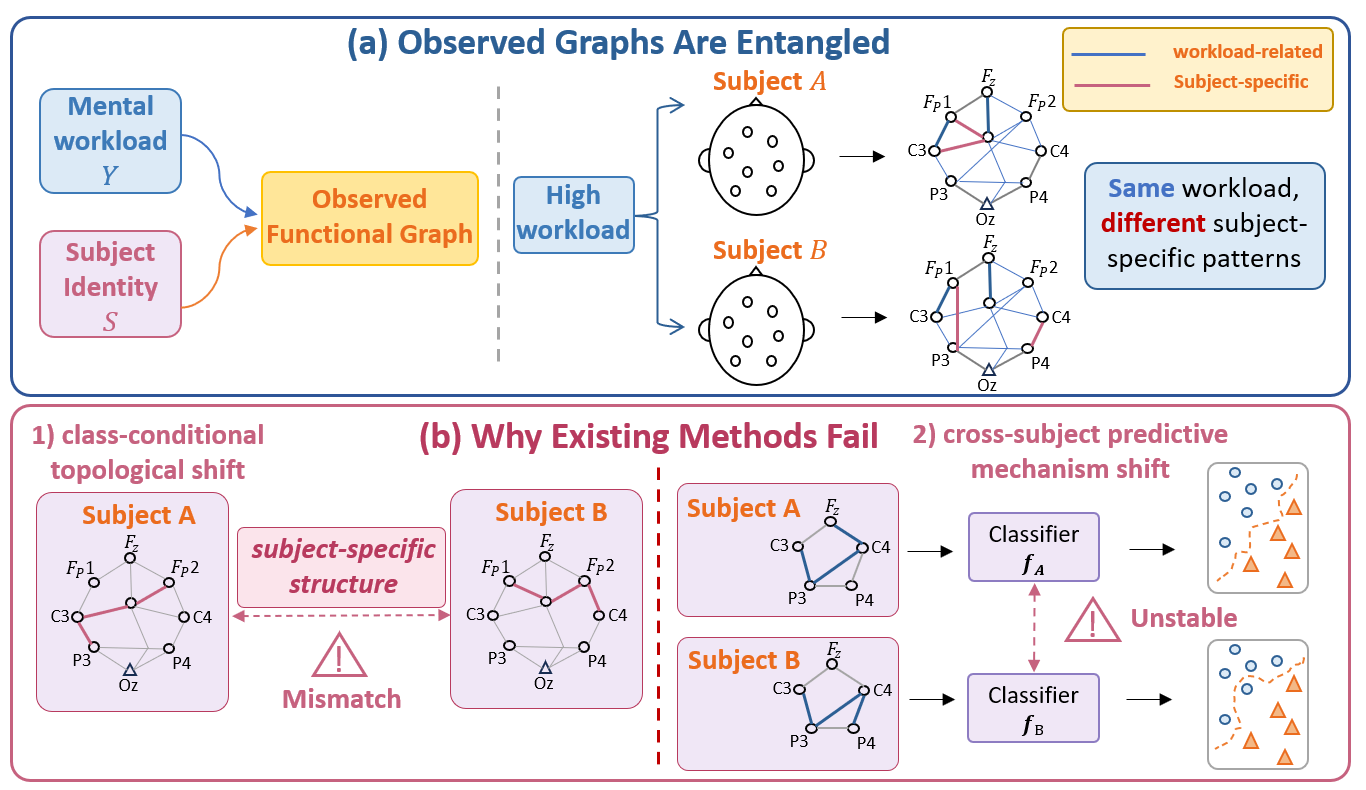} 
\caption{Two coupled shifts in cross-subject functional graph learning: class-conditional topological shift at the graph-rationale selection stage and cross-subject predictive mechanism shift at the classification stage.}
\label{fig:dual_shift}
\end{figure}

A promising approach to extract such task-relevant structures is graph rationale learning~\cite{chen2022ciga,miao2022gsat}, which identifies a compact, label-predictive subgraph prior to classification. However, simply optimizing empirical risk for subgraph selection in multi-subject EEG data is insufficient. Because subject-specific connectivity often acts as a spurious shortcut for source subjects, the model tends to exploit these variations, extracting rationales whose structures vary drastically across individuals. This causes a \emph{class-conditional topological shift} at the selection stage (Fig.~\ref{fig:dual_shift}): even under the same workload, the spectral properties of the extracted functional subgraphs differ significantly across subjects. Furthermore, even if a stable topology is extracted, the raw EEG node features traversing this topology inherently contain subject-specific baselines. This leads to a \emph{cross-subject predictive mechanism shift} at the classification stage, where the mapping from the learned graph representations to the final predictions becomes environment-dependent. While traditional cross-subject EEG methods attempt to align vector representations~\cite{albuquerque2022shifts,zhao2021plug,jeon2023mutual}, they largely overlook the upstream functional topology, thereby failing to resolve these two coupled shifts simultaneously.

To address these shifts, we propose \textbf{CDBG}, a unified \textbf{C}ausally motivated \textbf{D}ual-invariance learning framework for \textbf{B}rain \textbf{G}raphs. By treating each training subject as a distinct environment, CDBG regularizes graph rationale learning at both the topology and prediction levels. During rationale selection, a workload-conditional Laplacian spectral alignment mitigates topological shift by reducing class-conditional spectral discrepancies across environments. At the classification stage, subject-wise Invariant Risk Minimization (IRM)~\cite{arjovsky2019irm} promotes environment-wise risk stationarity, reducing sensitivity to predictive mechanism shift. Through joint optimization, the selected functional subgraphs are shaped by workload predictiveness, sparsity, and cross-subject stability. 

We evaluate CDBG on one self-built and two public EEG datasets under leave-one-subject-out (LOSO) protocols. Results show it significantly outperforms state-of-the-art baselines, improving Macro-F1 scores by 2.48\%, 3.73\%, and 4.23\% on SELF, STEW, and EEGMAT, respectively. Furthermore, comprehensive analyses validate the effectiveness of the proposed graph rationale selection and dual-invariance learning. The main contributions are summarized as follows:

\begin{itemize}
    \item We characterize cross-subject graph learning through two forms of distribution shift: class-conditional topological shift in the selected rationales and cross-subject predictive mechanism shift. This formulates subject-independent workload recognition as a dual-invariance problem over graph topology and prediction risk.
    \item We propose CDBG, a causally motivated graph rationale learning framework that extracts predictive connectivity via stochastic edge masking. It mitigates the dual shifts by combining workload-conditional Laplacian spectral alignment with subject-wise IRM regularization.

    \item We provide a theoretical bound connecting source spectral alignment to unseen-subject discrepancy, and a decomposition of the subject-wise IRM penalty. Extensive experiments under strict leave-one-subject-out protocols, alongside graph-level analyses, demonstrate the effectiveness and interpretability of the components.

\end{itemize}

\section{Related Work}

\paragraph{EEG-based Mental Workload Recognition.} 
Earlier workload and related cognitive-state pipelines relied on signal decomposition and hand-crafted features~\cite{yedukondalu2023cognitive,armanfard2016vigilance}. Deep models then enabled end-to-end spectral-temporal learning through convolution, recurrence, attention, and transfer learning~\cite{Wang_ARFN_2024,pulver2023feature,havugimana2023deep,lee2020continuous,xie2025cnn}. Motivated by functional connectivity, graph models explicitly represent EEG channels and their interactions. Early EEG GNNs learned adaptive adjacency or encoded biological topology~\cite{song2020dgcnn,zhong2022rgnn}, while later attention- and structure-learning variants spanned affective and clinical tasks~\cite{philipchen2025adamgraph,xiao2025dcgnn,ho2023anomalous}. Yet, these methods do not explicitly regularize workload-conditional topology across heterogeneous subject environments.

\paragraph{Cross-Subject Generalization and Invariant Graph Learning.}
Subject-to-subject distribution shift is a critical barrier for real-world EEG systems~\cite{albuquerque2022shifts}. While domain adaptation requires target-subject calibration data~\cite{zhao2021plug}, subject-independent generalization relies exclusively on source subjects. Existing approaches align multi-subject representations via adversarial inference~\cite{ozdenizci2020invariant}, mixture-of-experts~\cite{yang2025evomoe}, contrastive learning~\cite{chen2026emod,zhang2025cognitioncapturer,wu2026shrinking}, and large-scale self-supervised pretraining (e.g., EEGPT, State Mamba)~\cite{wang2024eegpt,Weng_StateMamba_2026}. However, these feature-level objectives do not directly regularize the structural invariance of functional connectivity. At the graph level, invariant rationale frameworks like CIGA~\cite{chen2022ciga} and GSAT~\cite{miao2022gsat} extract predictive subgraphs for general out-of-distribution scenarios. Yet, they are not specialized for subject-indexed functional brain graphs and fail to disentangle the highly coupled topology and node-attribute variations. CDBG addresses this by treating each subject as a distinct environment, uniquely targeting both the class-conditional topological shift and the predictive mechanism shift within a unified causally motivated framework.

\section{Method}
\label{sec:method}

\begin{figure*}[t]
    \centering
    \includegraphics[width=\textwidth]{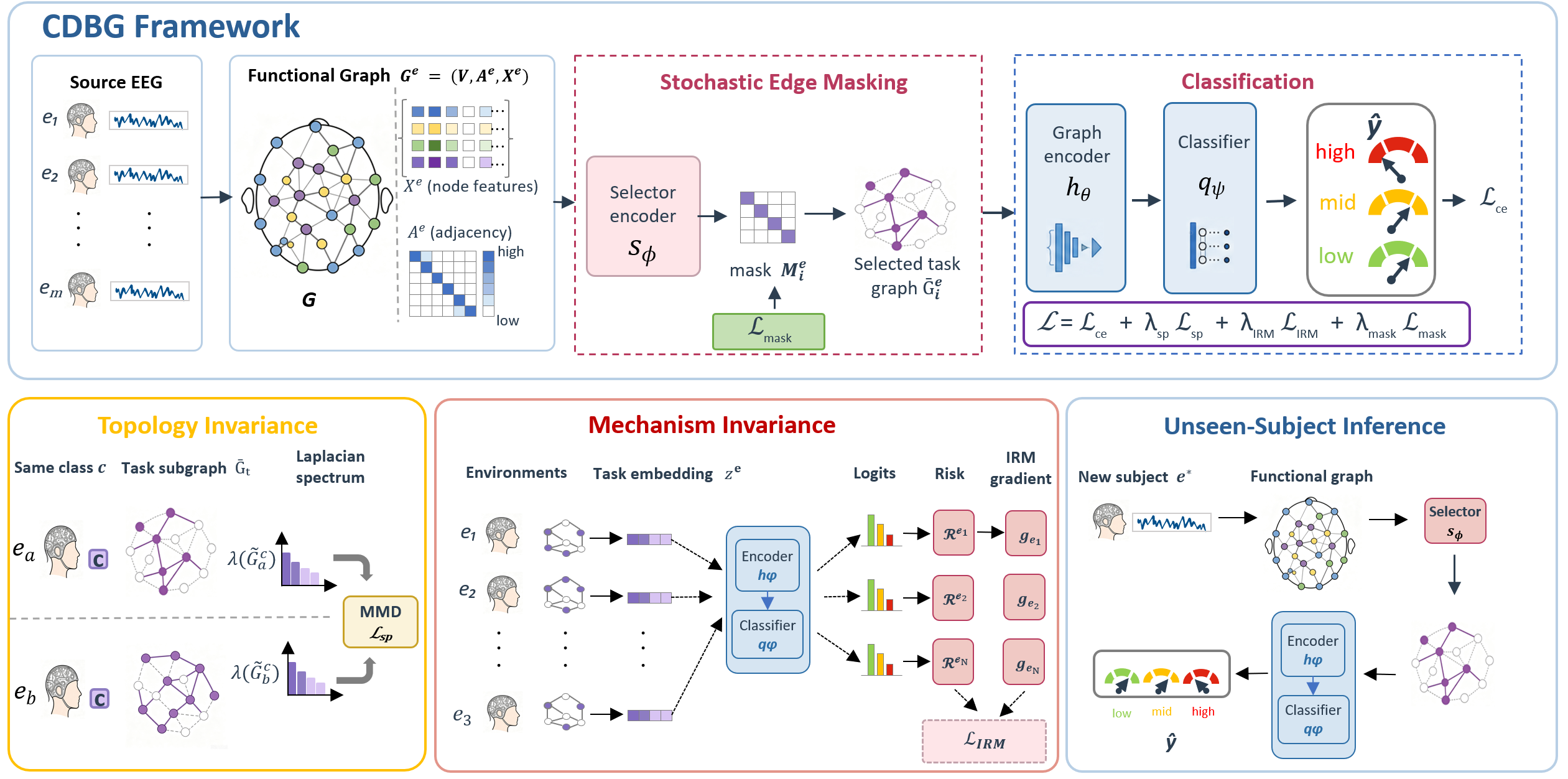}
    \caption{The overall framework of the proposed CDBG.}
    \label{fig:framework}
\end{figure*}

Motivated by the subject-modulated structure $Y \rightarrow G \leftarrow S$, we formulate subject-independent workload recognition as a causally motivated dual-invariance problem. Let each training subject represent a distinct environment $e \in \mathcal{E}_{\mathrm{tr}}$. As shown in Figure~\ref{fig:framework}, given an observed functional graph $G$, CDBG uses a learnable selector $s_\phi$ to extract a sparse rationale $\widetilde{G}$, and a graph encoder $h_\theta$ with a classifier $q_\psi$ maps $\widetilde{G}$ to the prediction $\hat{y}$.

To encourage cross-subject generalization, we impose two invariance constraints. At the selection stage, workload-conditional Laplacian spectral alignment mitigates \emph{class-conditional topological shift}. At the classification stage, subject-wise Invariant Risk Minimization (IRM) mitigates \emph{cross-subject predictive mechanism shift}.

\subsection{Problem Setup and Dual-Invariance Goals}

Let $\mathcal{D}^{e} = \{(G_i^{e},y_i^{e})\}_{i=1}^{n_e}$ denote the dataset from training subject $e\in\mathcal{E}_{\mathrm{tr}}$. Each EEG epoch is a graph $G_i^{e}=(V,A_i^{e},X_i^{e})$, with $C$ channels $V$, connectivity matrix $A_i^{e}\in\mathbb{R}^{C\times C}$, node features $X_i^{e}\in\mathbb{R}^{C\times F}$, and workload label $y_i^{e}\in\{1,\ldots,K\}$. Evaluated under the strict leave-one-subject-out (LOSO) protocol, the model targets an unseen environment $e^\ast\notin\mathcal{E}_{\mathrm{tr}}$.

Since $P^e(G,Y) = P(G,Y\mid S=e)$, the observed graph is subject-modulated. We regularize our pipeline $G \xrightarrow{s_\phi} \widetilde{G} \xrightarrow{h_\theta} Z \xrightarrow{q_\psi} \hat{Y}$ toward two ideal invariance conditions:

\textbf{1. Topological Invariance at Selection:} The class-conditional distribution of the extracted rationale $\widetilde{G}$ should remain stable across environments:
\begin{equation}
    P^e(\widetilde G\mid Y=c) \approx P^{e'}(\widetilde G\mid Y=c), \quad \forall e,e'\in\mathcal E_{\mathrm{tr}}.
    \label{eq:topology_invariance_goal}
\end{equation}

\textbf{2. Predictive Invariance at Classification:} The empirical relationship between the graph representation $Z = h_\theta(\widetilde{G})$ and label $Y$ should remain stable:
\begin{equation}
    P^e(Y\mid Z) \approx P^{e'}(Y\mid Z), \quad \forall e,e'\in\mathcal E_{\mathrm{tr}}.
    \label{eq:mechanism_invariance_goal}
\end{equation}

As enforcing Eq.~\ref{eq:topology_invariance_goal} and Eq.~\ref{eq:mechanism_invariance_goal} directly is intractable, we design tractable surrogate objectives.

\subsection{Mitigating Class-Conditional Topological Shift}

We employ a stochastic edge selector to extract a predictive subgraph and align its class-conditional spectral distributions across subjects.

\paragraph{Stochastic Edge Masking.}
To identify a task-relevant connectivity backbone, the selector $s_\phi$ uses an internal encoder to produce node representations $H^{(s)}$. For each candidate edge $(u,v)$, the unnormalized logit is $a_{uv,i}^e = (h_u^{(s)})^\top W_\phi h_v^{(s)}$, where $W_\phi$ is a learnable symmetric matrix. Using the Binary Concrete relaxation~\cite{maddison2017concrete}, we draw Logistic noise $\xi_{uv}$ and compute the continuous retention probability $m_{uv,i}^e \in (0,1)$ as:
\begin{equation}
    m_{uv,i}^e = \sigma\!\left( \frac{a_{uv,i}^e + \xi_{uv}}{\tau} \right),
\end{equation}
where $\sigma$ is the sigmoid and $\tau>0$ is the temperature. The soft adjacency matrix is $\widetilde A_i^e = A_i^e \odot M_i^e$, yielding the rationale $\widetilde G_i^e=(V,\widetilde A_i^e,X_i^e)$.

To control sparsity given a target retention ratio $r$, we penalize deviations from the expected edge density over a subject-balanced mini-batch $\mathcal B$:
\begin{equation}
    \mathcal{L}_{\mathrm{mask}} = \frac{1}{|\mathcal B|} \sum_{e\in\mathcal E_{\mathrm{tr}}} \sum_{i\in\mathcal B_e} \left| \frac{1}{|E_i^e|} \sum_{(u,v)\in E_i^e} m_{uv,i}^e - r \right|.
    \label{eq:mask_loss}
\end{equation}

\paragraph{Workload-Conditional Spectral Alignment.}
Optimizing for predictiveness alone cannot prevent the retained topology from overfitting to subject-specific structures. We utilize the normalized Laplacian spectrum to efficiently summarize macroscopic graph properties while filtering out high-frequency topological noise~\cite{chung1997spectral}. Treating $\widetilde A_i^e$ as weighted adjacency, the degree matrix is $[\widetilde D_i^e]_{uu} = \sum_v |[\widetilde A_i^e]_{uv}|$. The normalized Laplacian is $\widetilde L_i^e = I-(\widetilde D_i^e)^{-1/2} \widetilde A_i^e (\widetilde D_i^e)^{-1/2}$. We extract its first $d$ eigenvalues to form the spectral descriptor $\lambda(\widetilde G_i^e) = [\lambda_{i,1}^e,\ldots,\lambda_{i,d}^e] \in\mathbb{R}^d$.

To mitigate topological shift, we align these empirical descriptors for samples within the same workload class $c$. Let $\Lambda_e^c = \{\lambda(\widetilde G_i^e):i\in\mathcal B_e,\ y_i^e=c\}$. The topology-invariance loss utilizes Maximum Mean Discrepancy (MMD)~\cite{gretton2012mmd}:
\begin{equation}
    \mathcal{L}_{\mathrm{sp}} = \sum_{c:\,|\mathcal P_c|>0} \frac{1}{|\mathcal P_c|} \sum_{(e,e')\in\mathcal P_c} \mathrm{MMD}^2 \left(\Lambda_e^c,\Lambda_{e'}^c\right),
    \label{eq:spectral_loss}
\end{equation}
where $\mathcal P_c$ contains valid unique subject pairs for class $c$. As proved later, this explicit alignment bounds the unseen-subject spectral discrepancy.

\subsection{Mitigating Cross-Subject Predictive Shift}

Spectral alignment encourages structural consistency but overlooks the predictive variation induced by individual physiological baselines in the node features $X$~\cite{wang2024eegpt}. The representation $Z = h_\theta(\widetilde{G})$ must therefore support a shared decision rule whose risk is locally stationary across subjects.

Let $\ell_i^e=q_\psi(h_\theta(\widetilde G_i^e))$ denote the predicted logits. The standard empirical cross-entropy risk $\mathcal{L}_{\mathrm{ce}}$ alone easily exploits spurious subject associations. To approach Eq.~\ref{eq:mechanism_invariance_goal}, we adopt the Invariant Risk Minimization (IRMv1) penalty~\cite{arjovsky2019irm}. By introducing a dummy scalar $\rho$ to rescale the logits, the parameterized environment risk is $R^e(\phi,\theta,\psi,\rho) = \frac{1}{|\mathcal B_e|} \sum_{i\in\mathcal B_e} -\log [\operatorname{softmax}(\rho\,\ell_i^e)]_{y_i^e}$. We promote the shared scaling $\rho=1$ to be a first-order stationary point simultaneously across environments:
\begin{equation}
    \mathcal{L}_{\mathrm{IRM}} = \sum_{e\in\mathcal{E}_{\mathrm{tr}}} \left\| \nabla_{\rho}R^e(\phi,\theta,\psi,\rho) \big|_{\rho=1} \right\|_2^2.
    \label{eq:irm_loss}
\end{equation}
This penalty directly penalizes non-stationarity along the shared logit-scaling direction, serving as a practical surrogate for predictive invariance.

\subsection{Joint Optimization and Inference}

The overall objective of CDBG is optimized jointly end-to-end:
\begin{equation}
    \mathcal{L} = \mathcal{L}_{\mathrm{ce}} + \lambda_{\mathrm{sp}}\mathcal{L}_{\mathrm{sp}} + \lambda_{\mathrm{IRM}}\mathcal{L}_{\mathrm{IRM}} + \lambda_{\mathrm{mask}}\mathcal{L}_{\mathrm{mask}},
    \label{eq:overall_objective}
\end{equation}
where $\lambda$ hyperparameters balance the regularizers. Crucially, subject indices are required only during training. During inference, for an unseen target subject, the model executes the deterministic forward path $G\xrightarrow{s_\phi}\widetilde G\xrightarrow{h_\theta} Z \xrightarrow{q_\psi}\hat y$, making it highly efficient without test-time adaptation.

\section{Theoretical Analysis of Dual-Invariance}

We theoretically analyze how the proposed dual-invariance objectives encourage unseen-subject generalization. Specifically, we show that spectral alignment provides a conditional bound on target spectral discrepancy, while subject-wise IRM admits an exact cross-environment gradient variance decomposition.

\subsection{Target Spectral Discrepancy under Source Coverage}

For workload class $c$ and source $e$, let $P_e^c$ denote the population distribution of the spectral descriptor $\lambda(\widetilde G)$. Let $k$ be the characteristic kernel of MMD with RKHS $\mathcal H_k$, and kernel mean embedding $\mu_e^c = \mathbb E_{\lambda\sim P_e^c}[k(\lambda,\cdot)]$. For $m=|\mathcal E_{\mathrm{tr}}|$ source subjects, we define the uniform source mixture $\bar P^c=m^{-1}\sum_e P_e^c$ with mean embedding $\bar\mu^c=m^{-1}\sum_e\mu_e^c$, and the source spectral diameter $\Delta_c = \max_{e,e'} \operatorname{MMD}_k(P_e^c,P_{e'}^c)$.

\begin{assumption}[Source Coverage in Spectral RKHS]
\label{assum:source_coverage}
For each class $c$, the target subject's kernel mean embedding $\mu_*^c$ lies within the convex hull of the source embeddings: $\mu_*^c = \sum_{e\in\mathcal E_{\mathrm{tr}}} \alpha_e^c\mu_e^c$, where $\alpha_e^c\ge0$ and $\sum_e\alpha_e^c=1$.
\end{assumption}

This assumes the target spectral distribution is a convex mixture of source distributions, characterizing interpolation within source-observed variation rather than arbitrary extrapolation.

\begin{theorem}[Unseen-Subject Spectral Discrepancy]
\label{thm:target_spectral}
Under Assumption~\ref{assum:source_coverage}, for any class-specific spectral statistic $g_c\in\mathcal H_k$, the discrepancy between the target expectation and the source mixture expectation is bounded by the source spectral loss:
\begin{equation}
    \left| \mathbb E_{P_*^c}[g_c] - \mathbb E_{\bar P^c}[g_c] \right| \le \|g_c\|_{\mathcal H_k} \sqrt{|\mathcal P_c|\, \mathcal L_{\mathrm{sp},c}^{\mathrm{pop}}},
\end{equation}
where $\mathcal L_{\mathrm{sp},c}^{\mathrm{pop}}$ is the population counterpart of the pairwise loss in Eq.~\ref{eq:spectral_loss}.
\end{theorem}

\begin{corollary}[Topology-Dependent Spectral-Statistic Discrepancy]
\label{cor:target_spectral}
Let $\pi_c$ be the shared class prior. The target aggregate $T_{\mathrm{sp}}^* = \sum_c\pi_c\mathbb E_{P_*^c}[g_c]$ is bounded relative to the source mixture aggregate $\bar T_{\mathrm{sp}}$:
\begin{equation}
    |T_{\mathrm{sp}}^*-\bar T_{\mathrm{sp}}| \le \sum_c\pi_c \|g_c\|_{\mathcal H_k} \sqrt{|\mathcal P_c|\, \mathcal L_{\mathrm{sp},c}^{\mathrm{pop}}}.
\end{equation}
\end{corollary}

Theorem~\ref{thm:target_spectral} and Corollary~\ref{cor:target_spectral} show that aligning source spectral distributions provides a bound on unseen-subject spectral discrepancy, thus structurally regularizing topology-level shift. \textit{Detailed proofs are in the supplementary material.}

\begin{table*}[!t]
\centering
\footnotesize
\setlength{\tabcolsep}{2.5pt}
\renewcommand{\arraystretch}{0.9}
\begin{tabular}{llcccccc}
\toprule
Category & Method
& \multicolumn{2}{c}{SELF}
& \multicolumn{2}{c}{EEGMAT}
& \multicolumn{2}{c}{STEW}
\\
\cmidrule(lr){3-4} \cmidrule(lr){5-6} \cmidrule(lr){7-8}
& & ACC & F1 & ACC & F1 & ACC & F1 \\
\midrule

\multirow{4}{*}{Graph-Based}

& RGNN
& $21.74\,/\,7.54$ & $14.37\,/\,7.68$
& $64.37\,/\,12.83$ & $60.32\,/\,15.98$
& $73.21\,/\,14.87$ & $70.97\,/\,17.98$
\\

& DCGNN
& $24.81\,/\,6.21$ & $19.47\,/\,8.69$
& $63.38\,/\,13.26$ & $59.92\,/\,15.94$
& $71.00\,/\,13.02$ & $69.33\,/\,14.72$
\\

& AdamGraph
& $23.08\,/\,4.75$ & $17.16\,/\,6.97$
& $65.09\,/\,14.02$ & $61.83\,/\,16.33$
& $73.45\,/\,15.10$ & $71.78\,/\,17.05$
\\

& DGCNN
& $24.62\,/\,7.23$ & $19.79\,/\,8.29$
& $66.39\,/\,13.28$ & $62.23\,/\,17.17$
& $74.83\,/\,13.78$ & $72.92\,/\,16.49$
\\

\midrule

\multirow{4}{*}{Workload-Specific}

& BiConformer
& \underline{$25.87\,/\,6.97$} & $19.20\,/\,5.36$
& \underline{$70.14\,/\,13.49$} & \underline{$66.97\,/\,17.06$}
& $73.69\,/\,15.53$ & $72.16\,/\,17.57$
\\

& LSCCN
& $24.64\,/\,4.56$ & $20.58\,/\,5.10$
& $57.25\,/\,9.36$ & $55.22\,/\,10.49$
& $68.28\,/\,11.95$ & $66.88\,/\,13.56$
\\

& EEGMeNet
& $21.85\,/\,8.21$ & $14.93\,/\,7.70$
& $66.46\,/\,12.83$ & $61.97\,/\,17.18$
& $73.21\,/\,14.10$ & $71.14\,/\,17.17$
\\

& MuLHiTA
& $20.97\,/\,2.98$ & $17.07\,/\,4.34$
& $62.57\,/\,9.90$ & $60.32\,/\,11.48$
& $71.22\,/\,14.00$ & $70.01\,/\,15.22$
\\

\midrule

\multirow{4}{*}{Conventional EEG}

& EEG-Deformer
& $19.85\,/\,8.35$ & $15.86\,/\,6.07$
& $63.36\,/\,9.63$ & $59.46\,/\,12.66$
& $74.30\,/\,13.26$ & $72.18\,/\,16.24$
\\

& EEG-Conformer
& $22.01\,/\,6.16$ & $18.16\,/\,7.14$
& $65.30\,/\,14.04$ & $61.68\,/\,16.99$
& \underline{$75.51\,/\,10.78$} & \underline{$74.30\,/\,12.60$}
\\

& EEGLearn
& $24.87\,/\,13.48$ & \underline{$22.59\,/\,12.04$}
& $61.14\,/\,16.64$ & $56.23\,/\,20.20$
& $69.08\,/\,17.08$ & $63.41\,/\,23.04$
\\

& EEGNet
& $20.42\,/\,2.93$ & $14.55\,/\,3.77$
& $67.66\,/\,13.18$ & $65.35\,/\,15.65$
& $73.00\,/\,14.04$ & $71.09\,/\,16.29$
\\

\midrule

Ours
& \textbf{CDBG}
& \boldmath{$30.33\,/\,4.23$} & \boldmath{$25.07\,/\,3.30$}
& \boldmath{$73.08\,/\,11.96$} & \boldmath{$71.20\,/\,14.59$}
& \boldmath{$79.05\,/\,12.87$} & \boldmath{$78.03\,/\,14.75$}
\\

\bottomrule
\end{tabular}
\caption{Cross-subject comparison of CDBG and competing methods on three workload datasets. (mean $\,/\,$ standard deviation)}
\label{tab:cross}
\end{table*}

\subsection{Decomposition of the Subject-Wise IRM Penalty}

To analyze the geometric implication of the subject-wise IRM penalty in mitigating predictive shift, we define the environment-specific risk gradient $g_e = \nabla_\rho R^e(\phi,\theta,\psi,\rho) \big|_{\rho=1}$ and the average gradient $\bar g=\frac{1}{m}\sum_e g_e$.

\begin{proposition}[IRM Gradient Decomposition]
\label{prop:irm_decomposition}
The subject-wise IRM penalty admits an exact decomposition into average stationarity and cross-subject variance:
\begin{equation}
    \mathcal L_{\mathrm{IRM}} = m\|\bar g\|_2^2 + \sum_{e\in\mathcal E_{\mathrm{tr}}} \|g_e-\bar g\|_2^2.
    \label{eq:irm_decomposition}
\end{equation}
\end{proposition}

\begin{remark}[Pairwise Gradient Agreement]
\label{rem:gradient_agreement}
The aggregate pairwise gradient disagreement is upper-bounded by the IRM penalty: $\frac{1}{m} \sum_{e<e'}\|g_e-g_{e'}\|_2^2 \le \mathcal L_{\mathrm{IRM}}$.
\end{remark}

Proposition~\ref{prop:irm_decomposition} proves that the scalar IRM penalty practically enforces cross-subject prediction agreement by explicitly penalizing gradient variance. Jointly, these objectives regularize both topology-level shift and prediction-level risk non-stationarity.


%

\begin{table*}[!ht]
\centering
\footnotesize
\setlength{\tabcolsep}{2.5pt}
\renewcommand{\arraystretch}{0.9}

\begin{tabular}{llcccccc}
\toprule
Category & Method
& \multicolumn{2}{c}{SELF}
& \multicolumn{2}{c}{EEGMAT}
& \multicolumn{2}{c}{STEW}
\\
\cmidrule(lr){3-4}
\cmidrule(lr){5-6}
\cmidrule(lr){7-8}
& & ACC & F1 & ACC & F1 & ACC & F1
\\
\midrule

\multirow{4}{*}{Graph-Based}

& RGNN
& $3.91e-02$ & $1.95e-02$
& $2.66e-04$ & $2.07e-04$
& $3.08e-04$ & $5.05e-04$
\\

& DCGNN
& $2.73e-02$ & $9.77e-02$
& $9.29e-05$ & $2.55e-04$
& $4.77e-08$ & $1.51e-07$
\\

& AdamGraph
& $7.81e-03$ & $3.91e-02$
& $9.82e-05$ & $1.67e-04$
& $4.85e-04$ & $1.80e-04$
\\

& DGCNN
& $7.42e-02$ & $9.77e-02$
& $2.84e-03$ & $4.96e-04$
& $2.78e-03$ & $1.42e-03$
\\

\midrule

\multirow{4}{*}{Workload-Specific}

& BiConformer
& $1.91e-01$ & $5.47e-02$
& $1.13e-01$ & $5.94e-02$
& $1.01e-03$ & $4.85e-04$
\\

& LSCCN
& $7.42e-02$ & $1.25e-01$
& $8.93e-07$ & $1.65e-07$
& $1.02e-06$ & $4.13e-07$
\\

& EEGMeNet
& $1.95e-02$ & $1.17e-02$
& $4.21e-04$ & $1.08e-04$
& $6.39e-04$ & $1.22e-04$
\\

& MuLHiTA
& $3.91e-03$ & $3.91e-03$
& $1.94e-07$ & $5.40e-09$
& $3.18e-08$ & $4.40e-08$
\\

\midrule

\multirow{4}{*}{Conventional EEG}

& EEG-Deformer
& $3.91e-03$ & $3.91e-03$
& $1.46e-05$ & $8.82e-06$
& $1.73e-03$ & $4.85e-04$
\\

& EEG-Conformer
& $1.95e-02$ & $5.47e-02$
& $1.26e-04$ & $1.35e-04$
& $6.34e-03$ & $3.97e-03$
\\

& EEGLearn
& $1.91e-01$ & $2.30e-01$
& $2.60e-06$ & $2.44e-07$
& $3.52e-06$ & $1.77e-08$
\\

& EEGNet
& $3.91e-03$ & $3.91e-03$
& $7.12e-03$ & $1.60e-02$
& $4.21e-04$ & $1.07e-04$
\\

\bottomrule
\end{tabular}

\caption{Statistical significance analysis of CDBG under the cross-subject setting. Each entry reports the $p$-value from a paired Wilcoxon signed-rank test between CDBG and the corresponding baseline method.}
\label{tab:significance-cs}

\end{table*}

\section{Experiments}

\subsection{Experimental Settings}

This section details the benchmark setup used to evaluate CDBG, including the comparison methods, evaluation protocol, target datasets, and implementation configurations. Extended dataset statistics, baseline descriptions, dataset-specific settings, and comprehensive training parameters are provided in the supplementary material.

\subsubsection{Comparison Methods}

We compare CDBG with twelve representative baselines from three methodological categories. Graph-based models include RGNN~\cite{zhong2022rgnn}, DCGNN~\cite{xiao2025dcgnn}, AdamGraph~\cite{philipchen2025adamgraph}, and DGCNN~\cite{song2020dgcnn}. Workload-specific models include BiConformer~\cite{yan2025biconformer}, LSCCN~\cite{lsccn}, EEGMeNet~\cite{eegmenet}, and MuLHiTA~\cite{mulhita}. General EEG models include EEG-Deformer~\cite{ding2025eegdeformer}, EEG-Conformer~\cite{song2023eegconformer}, EEGLearn~\cite{eeglearn}, and EEGNet~\cite{lawhern2018eegnet}. Together, these methods cover convolutional, attention-based, and graph-based EEG representation learning.

\subsubsection{Evaluation Protocol}

CDBG is evaluated under a strict Leave-One-Subject-Out (LOSO) protocol. Since each subject is treated as an environment, the spectral alignment and IRM objectives require multiple training subjects and are not applicable to within-subject evaluation. In each fold, one subject is held out for testing, while the remaining subjects are used for training and model selection. We report Accuracy and Macro-F1, two widely used metrics in EEG workload recognition, with Macro-F1 as the primary metric. The reported results are averaged across all LOSO folds.

\subsubsection{Datasets}

To assess CDBG under diverse channel configurations and task complexities, we conduct experiments on three EEG mental workload datasets. SELF contains 59-channel EEG recordings collected at 500~Hz from 8 air traffic controllers under five simulated operational scenarios, covering rest, low, medium, high, and emergency workload conditions. EEGMAT~\cite{eegmat} contains 19-channel recordings from 36 subjects at 500~Hz under resting and mental arithmetic conditions. STEW~\cite{stew} contains 14-channel recordings from 48 subjects at 128~Hz under resting and multitasking conditions. EEGMAT and STEW provide established public benchmarks, whereas SELF introduces a professional air-traffic-control setting with more detailed workload levels and a more demanding classification task.

\subsubsection{Preprocessing and Implementation Details}

For each dataset, EEG signals are processed with dataset-specific bandpass filtering and a 50~Hz notch filter. The target signals are divided into non-overlapping 1-second segments at the original sampling rate. Channel-wise $z$-score normalization is then applied. All methods use the same processed inputs, LOSO partitions, and evaluation protocol. CDBG is optimized with Adam, and the checkpoint with the highest validation Macro-F1 is retained. The final hyperparameters are selected based on validation Macro-F1, with the tested ranges, chosen values, and related analysis reported in the Hyperparameter Analysis section and supplementary material. Its architecture is fixed across datasets, except for the temporal input size determined by the native sampling frequency. Baselines follow public implementations or published settings, with all tuning restricted to the validation data. The random number seed is set to 2026. Details of the computational setup used in the experiments are reported in the supplementary material.

\subsection{Experiments Results and Analysis}

\subsubsection{Comparison Experiments}

We compare CDBG with twelve baselines on SELF, STEW, and EEGMAT under the LOSO protocol. Performance is measured by ACC and Macro-F1, with percentage values reported as mean $\,/\,$ standard deviation. Table~\ref{tab:cross} summarizes the results, where the best and second-best values are marked in bold and underlined, respectively.

\noindent\textbf{Finding 1. CDBG achieves consistently superior performance across datasets.}
CDBG ranks first in both ACC and Macro-F1 on all three datasets. Compared with the strongest baselines, it improves Macro-F1 by 2.48, 4.23, and 3.73 percentage points on SELF, EEGMAT, and STEW, respectively. These results demonstrate its robustness across different workload settings and acquisition systems.

\noindent\textbf{Finding 2. CDBG remains effective in fine-grained workload recognition.}
SELF contains five workload states and only eight subjects, making cross-subject recognition particularly challenging. CDBG achieves an ACC of 30.33\% and a Macro-F1 of 25.07\%, surpassing the strongest baselines by 4.46 and 2.48 points, respectively. This suggests that CDBG better preserves discriminative workload information under limited subject diversity and fine-grained labels.

\noindent\textbf{Finding 3. Dual-invariance learning improves cross-subject generalization.}
CDBG consistently outperforms existing graph-based models. Compared with the strongest graph baseline, it improves Macro-F1 by 5.28, 8.97, and 5.11 points on SELF, EEGMAT, and STEW, respectively. These results indicate that modeling inter-channel dependencies alone is insufficient, while adaptive graph selection, spectral alignment, and invariant risk regularization jointly improve the transferability of workload representations.

We further conduct paired Wilcoxon signed-rank tests between CDBG and each baseline using subject-level ACC and Macro-F1 scores. As shown in Table~\ref{tab:significance-cs}, more than 80\% of the comparisons are statistically significant, indicating that the improvements are consistent across test subjects rather than driven by a small number of cases.

\subsubsection{Ablation Study}

We assess the contributions of spectral alignment $\mathcal{L}_{\mathrm{sp}}$, invariant risk minimization $\mathcal{L}_{\mathrm{IRM}}$, and adaptive graph masking $\mathcal{L}_{\mathrm{mask}}$ by removing each component individually while keeping the backbone and training settings unchanged. Table~\ref{tab:cdbg-ablation-cs} reports ACC and Macro-F1 as mean $\,/\,$ standard deviation.

The results confirm the complementary roles of the three components. First, removing $\mathcal{L}_{\mathrm{sp}}$ consistently reduces performance, showing that spectral alignment helps preserve stable graph structures. Second, removing $\mathcal{L}_{\mathrm{IRM}}$ produces the largest Macro-F1 reductions on SELF and EEGMAT, supporting its role in learning invariant workload representations. Third, removing $\mathcal{L}_{\mathrm{mask}}$ causes consistent degradation, indicating that connectivity selection retains informative edges while suppressing irrelevant structures.

\begin{figure*}[!t]
\centering
\begin{subfigure}[t]{0.32\textwidth}
    \centering
    \includegraphics[width=\linewidth]{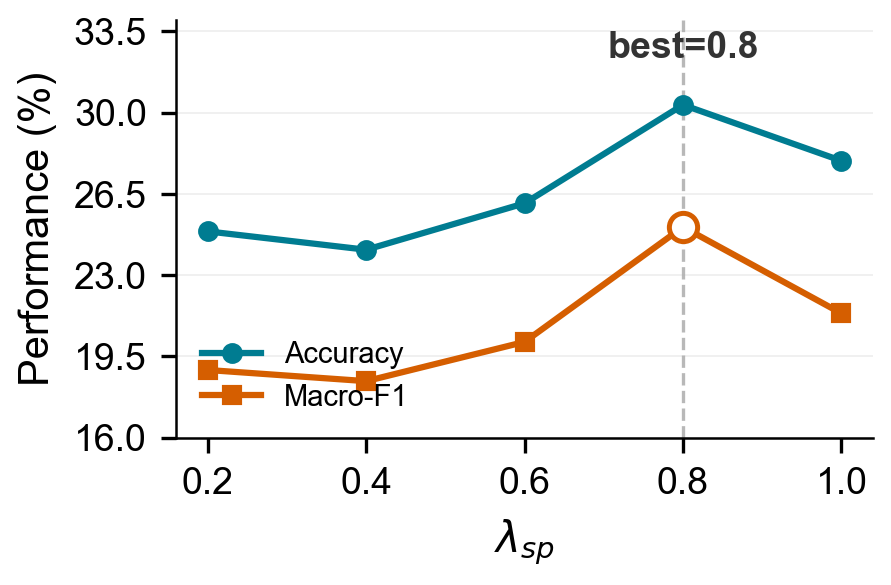}
    \caption{Spectral alignment weight $\lambda_{\mathrm{sp}}$.}
    \label{fig:hyperparameter-sp}
\end{subfigure}
\hfill
\begin{subfigure}[t]{0.32\textwidth}
    \centering
    \includegraphics[width=\linewidth]{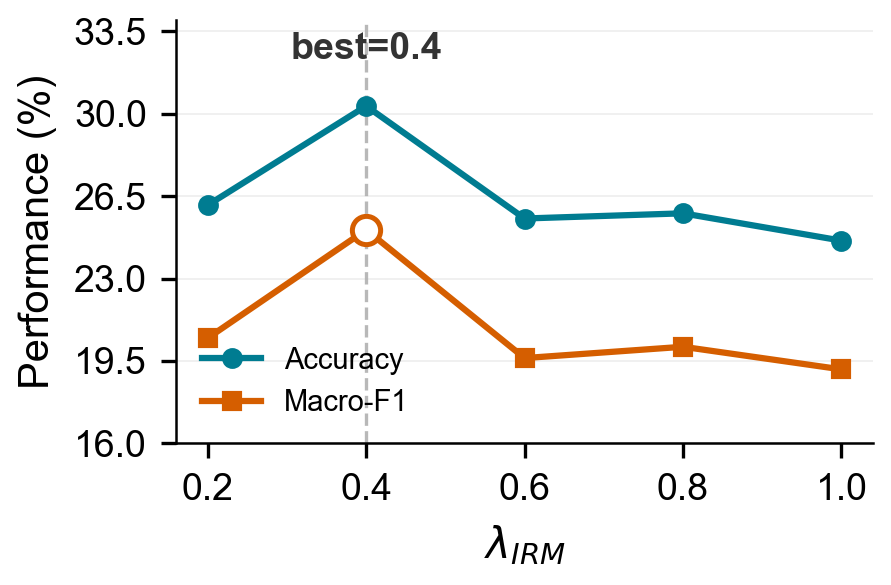}
    \caption{Invariant risk weight $\lambda_{\mathrm{IRM}}$.}
    \label{fig:hyperparameter-irm}
\end{subfigure}
\hfill
\begin{subfigure}[t]{0.32\textwidth}
    \centering
    \includegraphics[width=\linewidth]{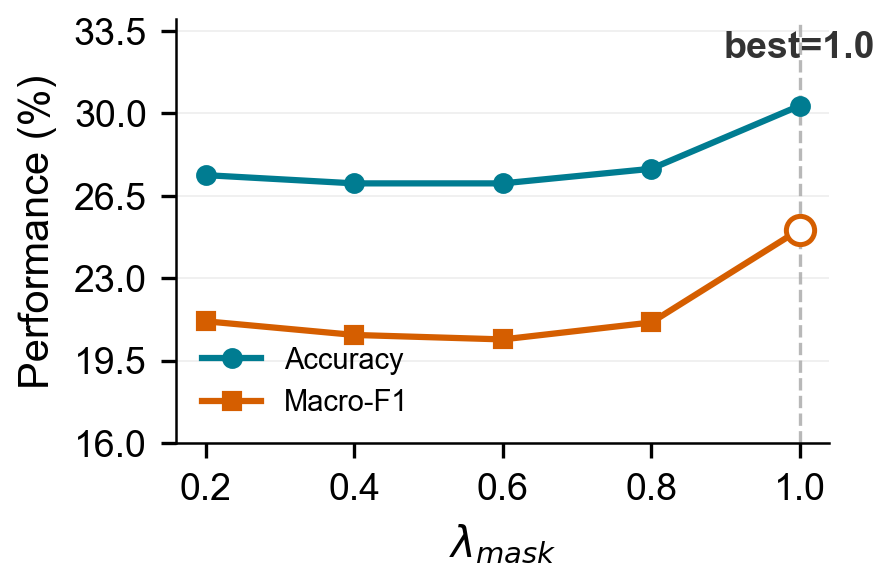}
    \caption{Mask regularization weight $\lambda_{\mathrm{mask}}$.}
    \label{fig:hyperparameter-mask}
\end{subfigure}
\caption{Hyperparameter analysis of CDBG on SELF under cross-subject evaluation.}
\label{fig:hyperparameter-analysis}
\end{figure*}

\begin{table}[!t]
\centering
\scriptsize
\setlength{\tabcolsep}{1.2pt}
\renewcommand{\arraystretch}{1.06}
\resizebox{\columnwidth}{!}{
\begin{tabular}{llcccc}
\toprule
Model & Metric
& Full
& w/o $\mathcal{L}_{\mathrm{sp}}$
& w/o $\mathcal{L}_{\mathrm{IRM}}$
& w/o $\mathcal{L}_{\mathrm{mask}}$
\\
\midrule

\multirow{2}{*}{SELF}
& ACC
& $\boldsymbol{30.33\,/\,4.23}$
& $26.76\,/\,2.89$
& $25.92\,/\,2.86$
& $26.15\,/\,4.97$
\\
& F1
& $\boldsymbol{25.07\,/\,3.30}$
& $20.70\,/\,3.12$
& $20.10\,/\,3.55$
& $21.18\,/\,4.31$
\\

\midrule

\multirow{2}{*}{STEW}
& ACC
& $\boldsymbol{79.05\,/\,12.87}$
& $74.76\,/\,11.46$
& $74.91\,/\,11.50$
& $74.64\,/\,11.53$
\\
& F1
& $\boldsymbol{78.03\,/\,14.75}$
& $73.45\,/\,13.24$
& $73.66\,/\,13.27$
& $73.41\,/\,13.17$
\\

\midrule

\multirow{2}{*}{EEGMAT}
& ACC
& $\boldsymbol{73.08\,/\,11.96}$
& $64.85\,/\,11.21$
& $64.29\,/\,10.84$
& $64.72\,/\,10.73$
\\
& F1
& $\boldsymbol{71.20\,/\,14.59}$
& $62.30\,/\,13.38$
& $61.49\,/\,12.78$
& $61.83\,/\,13.12$
\\

\bottomrule
\end{tabular}
}
\caption{Cross-subject ablation results of CDBG.}
\label{tab:cdbg-ablation-cs}
\end{table}

\begin{figure}[!t]
\centering
\includegraphics[width=0.9\columnwidth]{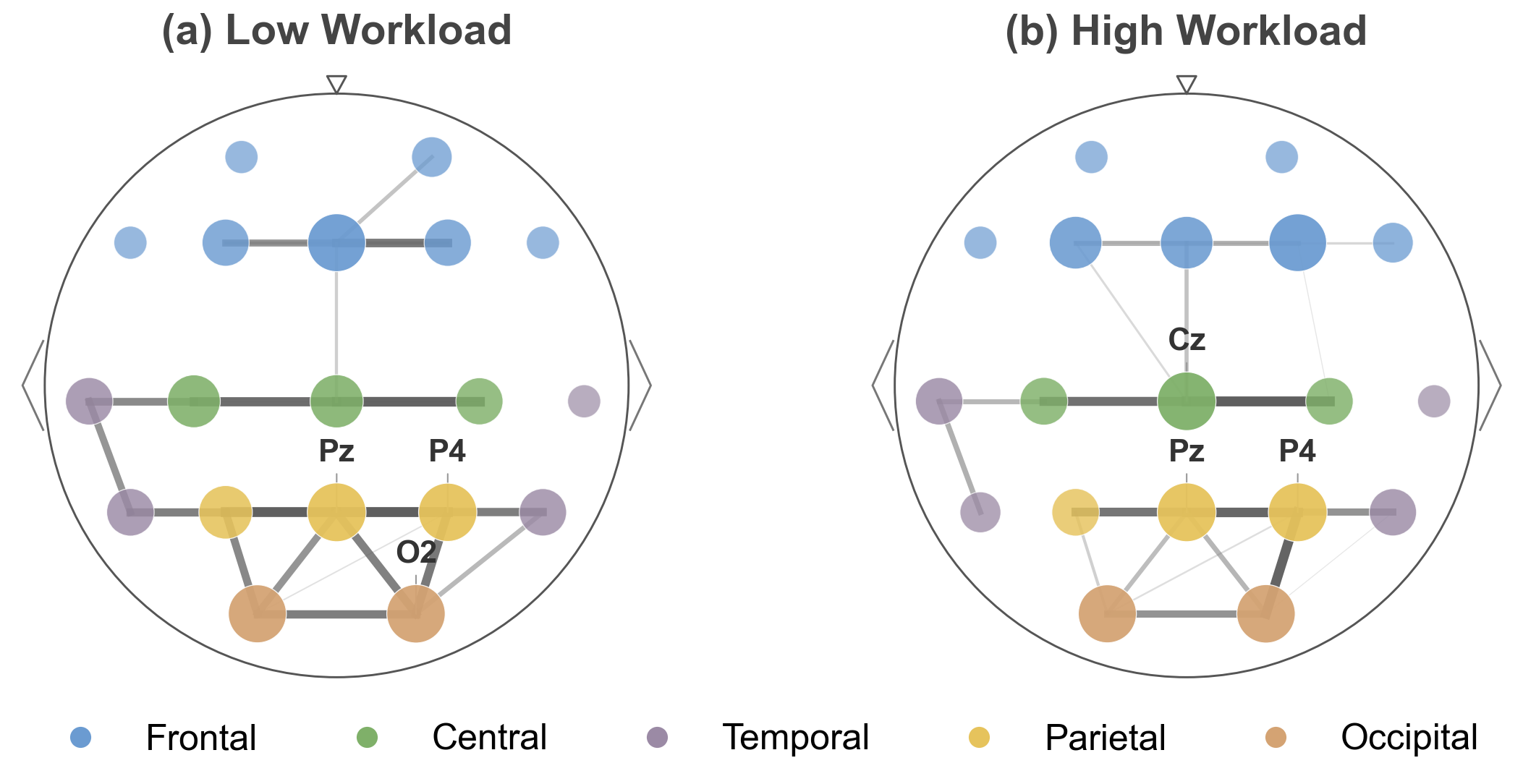}
\caption{Workload-dependent functional connectivity patterns and channel contributions on EEGMAT.}
\label{fig:cdbg-connectivity}
\end{figure}

\subsubsection{Hyperparameter Analysis}

We examine the three regularization coefficients in the joint objective, including the spectral alignment weight $\lambda_{\mathrm{sp}}$, the invariant risk weight $\lambda_{\mathrm{IRM}}$, and the edge-mask regularization weight $\lambda_{\mathrm{mask}}$. For each coefficient, we search over $\{0.2,0.4,0.6,0.8,1.0\}$ while fixing the other settings. Experiments are conducted on SELF under cross-subject evaluation, and the results are presented in Fig.~\ref{fig:hyperparameter-analysis}.

The best performance is achieved with $\lambda_{\text{sp}} = 0.8$, $\lambda_{\text{IRM}} = 0.4$, and $\lambda_{\text{mask}} = 1.0$. A moderate $\lambda_{\text{sp}}$ balances class-conditional topology alignment with the preservation of discriminative structures. The lower optimal value for $\lambda_{\text{IRM}}$ suggests that encouraging predictive invariance is beneficial, though an overemphasized gradient penalty degrades performance. Increasing $\lambda_{\text{mask}}$ to $1.0$ effectively controls graph density and suppresses redundant edges, whereas smaller values provide insufficient sparsity regularization. Overall, performance exhibits single-peak trends across the tested ranges, confirming the importance of appropriate weighting and each loss component in CDBG.

\subsubsection{Workload-Dependent Connectivity Interpretation}

We inspect the functional structures retained by CDBG on EEGMAT under cross-subject evaluation. For each workload condition, Fig.~\ref{fig:cdbg-connectivity} displays the selected connectivity pattern, where edge width reflects connection strength and node size summarizes the contribution of incident connections.

The learned graphs reveal a clear workload-dependent reorganization. Under low workload, prominent connections occur within the frontal, central, and parieto-occipital regions. Under high workload, Cz, Pz, P4, and O2 become more prominent, accompanied by stronger cross-regional integration among central, parietal, and occipital channels. This shift is consistent with the increased attentional coordination and neural integration required by the SIMKAP multitasking condition. The distinct topologies therefore suggest that CDBG captures workload-sensitive connectivity patterns with plausible neurophysiological interpretations.

\begin{figure}[!t]
\centering
\includegraphics[width=0.8\columnwidth]{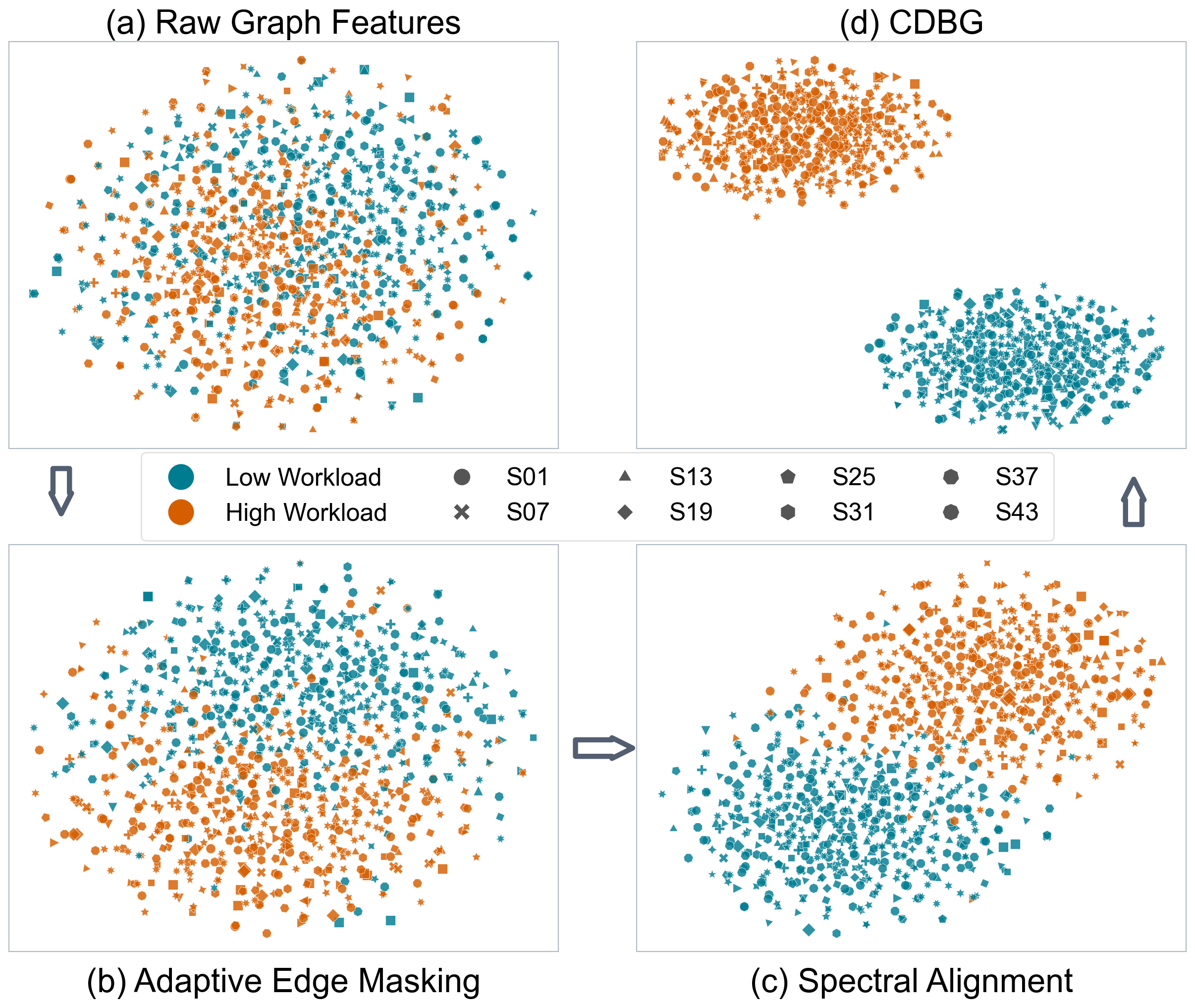}
\caption{t-SNE visualization of learned representations on STEW under progressively constrained model settings.}
\label{fig:cdbg-tsne}
\end{figure}

\subsubsection{Cross-Subject Representation Analysis}

To examine how the proposed components reshape the representation space, we apply t-SNE to held-out STEW samples from four progressively constrained settings: raw graph features, adaptive edge masking, masking with workload-conditional spectral alignment, and full CDBG with IRM. As shown in Fig.~\ref{fig:cdbg-tsne}, raw representations exhibit substantial class overlap and subject-dependent grouping. Adaptive masking removes irrelevant connectivity, while spectral alignment brings same-class samples from different subjects closer. IRM further separates workload classes, including samples from the same subject. Consequently, the full model yields the clearest class boundaries and weakest subject-related clustering, providing qualitative evidence for the complementary effects of topology selection, spectral invariance, and predictive invariance.

\section{Conclusion}

In this paper, we study cross-subject EEG-based mental workload recognition from a causal perspective. We identify two coupled shifts caused by inter-subject variability: class-conditional topological shift in functional connectivity and cross-subject predictive mechanism shift. To address them, we propose CDBG, a causal dual-invariance framework. By jointly optimizing stochastic graph rationale selection, workload-conditional Laplacian spectral alignment, and subject-wise Invariant Risk Minimization, CDBG separates task-intrinsic neural signatures from subject-specific variations. Experiments under strict leave-one-subject-out protocols show that CDBG achieves state-of-the-art performance while yielding neurophysiologically interpretable functional rationales. Future work will extend the framework to other EEG-based cognitive tasks and explore unsupervised test-time adaptation for real-world deployment.

\bibliography{aaai2027}

@article{eegmat,
  AUTHOR = {Zyma, Igor and Tukaev, Sergii and Seleznov, Ivan and Kiyono, Ken and Popov, Anton and Chernykh, Mariia and Shpenkov, Oleksii},
  TITLE = {Electroencephalograms during Mental Arithmetic Task Performance},
  JOURNAL = {Data},
  VOLUME = {4},
  YEAR = {2019},
  NUMBER = {1},
  ARTICLE-NUMBER = {14},
  URL = {https://www.mdpi.com/2306-5729/4/1/14},
  ISSN = {2306-5729},
  DOI = {10.3390/data4010014}
}

@article{stew,
  author={Lim, W. L. and Sourina, O. and Wang, L. P.},
  journal={IEEE Transactions on Neural Systems and Rehabilitation Engineering}, 
  title={STEW: Simultaneous Task EEG Workload Data Set}, 
  year={2018},
  volume={26},
  number={11},
  pages={2106-2114},
  doi={10.1109/TNSRE.2018.2872924}
}

@article{lsccn,
  author={Yu, Yinhu and Bezerianos, Anastasios and Cichocki, Andrzej and Li, Junhua},
  journal={IEEE Transactions on Neural Systems and Rehabilitation Engineering}, 
  title={Latent Space Coding Capsule Network for Mental Workload Classification}, 
  year={2023},
  volume={31},
  number={},
  pages={3417-3427},
  doi={10.1109/TNSRE.2023.3307481}
}

@article{mulhita,
  author={Xia, Likun and Feng, Yuan and Guo, Ziheng and Ding, Jinhong and Li, Yanlei and Li, Yifan and Ma, Ming and Gan, Guoxi and Xu, Yehan and Luo, Jingyu and Shi, Zhiping and Guan, Yong},
  journal={IEEE Transactions on Neural Networks and Learning Systems}, 
  title={MuLHiTA: A Novel Multiclass Classification Framework With Multibranch LSTM and Hierarchical Temporal Attention for Early Detection of Mental Stress}, 
  year={2023},
  volume={34},
  number={12},
  pages={9657-9670},
  doi={10.1109/TNNLS.2022.3159573}
}

@article{eegmenet,
  author={Kongwudhikunakorn, Supavit and Ponwitayarat, Wuttikorn and Kiatthaveephong, Suktipol and Polpakdee, Wipamas and Yagi, Tohru and Senanarong, Vorapun and Ittichaiwong, Piyalitt and Wilaiprasitporn, Theerawit},
  journal={IEEE Internet of Things Journal}, 
  title={EEGMeNet: End-to-End Multitask Neural Network for Brain-Based Mental Workload Classification}, 
  year={2025},
  volume={12},
  number={20},
  pages={42573-42589},
  doi={10.1109/JIOT.2025.3593907}
}

@article{yan2025biconformer,
  author={Yan, Xu and Shi, Yuhang and Zhang, Tingting and Chen, Xin and Mao, Yi},
  journal={IEEE Transactions on Instrumentation and Measurement}, 
  title={BiConformer: EEG-Based Cognitive Load Assessment Using a Bidirectional Convolutional Transformer Network}, 
  year={2025},
  volume={74},
  number={},
  pages={1-15},
  doi={10.1109/TIM.2025.3633358}
}

@inproceedings{eeglearn,
  author    = {Bashivan, Pouya and Rish, Irina and Yeasin, Mohammed and Codella, Noel},
  title     = {Learning Representations from {EEG} with Deep Recurrent-Convolutional Neural Networks},
  booktitle = {International Conference on Learning Representations (ICLR)},
  year      = {2016}
}

@article{lawhern2018eegnet,
   title={EEGNet: a compact convolutional neural network for EEG-based brain–computer interfaces},
   volume={15},
   ISSN={1741-2552},
   url={http://dx.doi.org/10.1088/1741-2552/aace8c},
   DOI={10.1088/1741-2552/aace8c},
   number={5},
   journal={Journal of Neural Engineering},
   publisher={IOP Publishing},
   author={Lawhern, Vernon J and Solon, Amelia J and Waytowich, Nicholas R and Gordon, Stephen M and Hung, Chou P and Lance, Brent J},
   year={2018},
   pages={056013} 
}

@article{song2023eegconformer,
  author={Song, Yonghao and Zheng, Qingqing and Liu, Bingchuan and Gao, Xiaorong},
  journal={IEEE Transactions on Neural Systems and Rehabilitation Engineering}, 
  title={EEG Conformer: Convolutional Transformer for EEG Decoding and Visualization}, 
  year={2023},
  volume={31},
  number={},
  pages={710-719},
  doi={10.1109/TNSRE.2022.3230250}
}

@article{ding2025eegdeformer,
  author={Ding, Yi and Li, Yong and Sun, Hao and Liu, Rui and Tong, Chengxuan and Liu, Chenyu and Zhou, Xinliang and Guan, Cuntai},
  journal={IEEE Journal of Biomedical and Health Informatics}, 
  title={EEG-Deformer: A Dense Convolutional Transformer for Brain-Computer Interfaces}, 
  year={2025},
  volume={29},
  number={3},
  pages={1909-1918},
  doi={10.1109/JBHI.2024.3504604}
}

@article{song2020dgcnn,
  author={Song, Tengfei and Zheng, Wenming and Song, Peng and Cui, Zhen},
  journal={IEEE Transactions on Affective Computing}, 
  title={EEG Emotion Recognition Using Dynamical Graph Convolutional Neural Networks}, 
  year={2020},
  volume={11},
  number={3},
  pages={532-541},
  doi={10.1109/TAFFC.2018.2817622}
}

@article{zhong2022rgnn,
  author={Zhong, Peixiang and Wang, Di and Miao, Chunyan},
  journal={IEEE Transactions on Affective Computing}, 
  title={EEG-Based Emotion Recognition Using Regularized Graph Neural Networks}, 
  year={2022},
  volume={13},
  number={3},
  pages={1290-1301},
  doi={10.1109/TAFFC.2020.2994159}
}

@article{philipchen2025adamgraph,
  author={Philip Chen, C. L. and Chen, Bianna and Zhang, Tong},
  journal={IEEE Transactions on Cybernetics}, 
  title={AdamGraph: Adaptive Attention-Modulated Graph Network for EEG Emotion Recognition}, 
  year={2025},
  volume={55},
  number={5},
  pages={2038-2051},
  doi={10.1109/TCYB.2025.3550191}
}

@article{xiao2025dcgnn,
  author={Xiao, Yushun and Zheng, Wenming and Zhao, Guoying},
  journal={IEEE Transactions on Affective Computing}, 
  title={Dynamical Causal Graph Neural Network for EEG Emotion Recognition}, 
  year={2025},
  volume={16},
  number={4},
  pages={2803-2815},
  doi={10.1109/TAFFC.2025.3582740}
}

@article{yedukondalu2023cognitive,
  title={Cognitive Load Detection Using Circulant Singular Spectrum Analysis and Binary Harris Hawks Optimization Based Feature Selection},
  author={Yedukondalu, Jammisetty and Sharma, Lakhan Dev},
  journal={Biomedical Signal Processing and Control},
  volume={79},
  pages={104006},
  year={2023},
  doi={10.1016/j.bspc.2022.104006}
}

@misc{havugimana2023deep,
  title={Deep Learning Framework for Modeling Cognitive Load from Small and Noisy {EEG} Data},
  author={Havugimana, Felix and Moinuddin, Kazi and Yeasin, M.},
  year={2023},
  howpublished={TechRxiv preprint},
  doi={10.36227/techrxiv.21637409.v2}
}

@inproceedings{pulver2023feature,
  title={{EEG}-Based Cognitive Load Classification Using Feature Masked Autoencoding and Emotion Transfer Learning},
  author={Pulver, Dustin and Angkan, Prithila and Hungler, Paul and Etemad, Ali},
  booktitle={Proceedings of the 25th International Conference on Multimodal Interaction},
  pages={190--197},
  year={2023},
  publisher={Association for Computing Machinery},
  doi={10.1145/3577190.3614113}
}

@article{lee2020continuous,
  title={Continuous {EEG} Decoding of Pilots' Mental States Using Multiple Feature Block-Based Convolutional Neural Network},
  author={Lee, Dae-Hyeok and Jeong, Ji-Hoon and Kim, Kiduk and Yu, Baek-Woon and Lee, Seong-Whan},
  journal={IEEE Access},
  volume={8},
  pages={121929--121941},
  year={2020},
  doi={10.1109/ACCESS.2020.3006907}
}

@inproceedings{armanfard2016vigilance,
  title={Vigilance Lapse Identification Using Sparse {EEG} Electrode Arrays},
  author={Armanfard, Narges and Komeili, Majid and Reilly, James P. and Pino, Lou},
  booktitle={2016 IEEE Canadian Conference on Electrical and Computer Engineering},
  pages={1--4},
  year={2016},
  doi={10.1109/CCECE.2016.7726846}
}

@article{xie2025cnn,
  title={{CNN}-Transformer Network for Student Learning Effect Prediction Using {EEG} Signals Based on Spatio-Temporal Feature Fusion},
  author={Xie, Hui and Dong, Zexiao and Yang, Huiting and Luo, Yanxia and Ren, Shenghan and Zhang, Pengyuan and He, Jiangshan and Jia, Chunli and Yang, Yuqiang and Jiang, Mingzhe and Gao, Xinbo and Chen, Xueli},
  journal={Applied Soft Computing},
  volume={170},
  pages={112631},
  year={2025},
  doi={10.1016/j.asoc.2024.112631}
}

@article{ho2023anomalous,
  title={Self-Supervised Learning for Anomalous Channel Detection in {EEG} Graphs: Application to Seizure Analysis},
  author={Ho, Thi Kieu Khanh and Armanfard, Narges},
  journal={Proceedings of the AAAI Conference on Artificial Intelligence},
  volume={37},
  number={7},
  pages={7866--7874},
  year={2023},
  doi={10.1609/aaai.v37i7.25952}
}

@article{ozdenizci2020invariant,
  title={Learning Invariant Representations from {EEG} via Adversarial Inference},
  author={{\"O}zdenizci, Ozan and Wang, Ye and Koike-Akino, Toshiaki and Erdo{\u{g}}mu{\c{s}}, Deniz},
  journal={IEEE Access},
  volume={8},
  pages={27074--27085},
  year={2020},
  doi={10.1109/ACCESS.2020.2971600}
}

@article{yang2025evomoe,
  title={{EvoMoE}: Evolutionary Mixture-of-Experts for {SSVEP-EEG} Classification with User-Independent Training},
  author={Yang, Xiaoli and Li, Yurui and Zhang, Jianyu and Tian, Huiyuan and Li, Shijian and Pan, Gang},
  journal={IEEE Journal of Biomedical and Health Informatics},
  volume={29},
  number={9},
  pages={6538--6550},
  year={2025},
  doi={10.1109/JBHI.2025.3565882}
}

@article{chen2026emod,
  title={{EMOD}: A Unified {EEG} Emotion Representation Framework Leveraging {V-A} Guided Contrastive Learning},
  author={Chen, Yuning and Zhao, Sha and Li, Shijian and Pan, Gang},
  journal={Proceedings of the AAAI Conference on Artificial Intelligence},
  volume={40},
  number={21},
  pages={17427--17435},
  year={2026},
  doi={10.1609/aaai.v40i21.38796}
}

@article{zhang2025cognitioncapturer,
  title={CognitionCapturer: Decoding Visual Stimuli from Human {EEG} Signal with Multimodal Information},
  author={Zhang, Kaifan and He, Lihuo and Jiang, Xin and Lu, Wen and Wang, Di and Gao, Xinbo},
  journal={Proceedings of the AAAI Conference on Artificial Intelligence},
  volume={39},
  number={13},
  pages={14486--14493},
  year={2025},
  doi={10.1609/aaai.v39i13.33587}
}

@article{wu2026shrinking,
  title={Shrinking the Teacher: An Adaptive Teaching Paradigm for Asymmetric {EEG}-Vision Alignment},
  author={Wu, Lukun and Li, Jie and Ren, Ziqi and Zhang, Kaifan and Gao, Xinbo},
  journal={Proceedings of the AAAI Conference on Artificial Intelligence},
  volume={40},
  number={21},
  pages={17859--17867},
  year={2026},
  doi={10.1609/aaai.v40i21.38844}
}

@article{lotte2018review,
  title={A review of classification algorithms for EEG-based brain-computer interfaces: A 10 year update},
  author={Lotte, Fabien and Bougrain, Laurent and Cichocki, Andrzej and Clerc, Maureen and Congedo, Marco and Rakotomamonjy, Alain and Yger, Florian},
  journal={Journal of Neural Engineering},
  volume={15},
  number={3},
  pages={031005},
  year={2018},
  doi={10.1088/1741-2552/aab2f2}
}

@inproceedings{wang2024eegpt,
 author = {Wang, Guagnyu and Liu, Wenchao and He, Yuhong and Xu, Cong and Ma, Lin and Li, Haifeng},
 booktitle = {Advances in Neural Information Processing Systems},
 doi = {10.52202/079017-1239},
 pages = {39249--39280},
 publisher = {Curran Associates, Inc.},
 title = {EEGPT: Pretrained Transformer for Universal and Reliable Representation of EEG Signals},
 volume = {37},
 year = {2024}
}

@inproceedings{arjovsky2019irm,
  title={Invariant Risk Minimization},
  author={Arjovsky, Martin and Bottou, Leon and Gulrajani, Ishaan and Lopez-Paz, David},
  booktitle={arXiv preprint arXiv:1907.02893},
  year={2019}
}

@inproceedings{chen2022ciga,
  title={Learning Causally Invariant Representations for Out-of-Distribution Generalization on Graphs},
  author={Chen, Yongqiang and Zhang, Yonggang and Bian, Yatao and Yang, Han and Ma, Kaili and Xie, Binghui and Liu, Tongliang and Han, Bo and Cheng, James},
  booktitle={Advances in Neural Information Processing Systems},
  volume={35},
  pages={22131--22148},
  year={2022},
  doi={10.52202/068431-1608}
}

@article{gretton2012mmd,
  title={A Kernel Two-Sample Test},
  author={Gretton, Arthur and Borgwardt, Karsten M. and Rasch, Malte J. and Scholkopf, Bernhard and Smola, Alexander},
  journal={Journal of Machine Learning Research},
  volume={13},
  pages={723--773},
  year={2012}
}

@article{albuquerque2022shifts,
  title={Estimating distribution shifts for predicting cross-subject generalization in electroencephalography-based mental workload assessment},
  author={Albuquerque, Isabela and Monteiro, Jo{\~a}o and Rosanne, Olivier and Falk, Tiago H.},
  journal={Frontiers in Artificial Intelligence},
  volume={5},
  pages={992732},
  year={2022},
  doi={10.3389/frai.2022.992732}
}

@article{klepl2024survey,
  title={Graph neural network-based eeg classification: A survey},
  author={Klepl, Dominik and Wu, Min and He, Fei},
  journal={IEEE Transactions on Neural Systems and Rehabilitation Engineering},
  volume={32},
  pages={493--503},
  year={2024},
  publisher={IEEE}
}

@article{safari2024classification,
  title   = {Classification of Mental Workload Using Brain Connectivity and Machine Learning on Electroencephalogram Data},
  author  = {Safari, MohammadReza and Shalbaf, Reza and Bagherzadeh, Sara and Shalbaf, Ahmad},
  journal = {Scientific Reports},
  volume  = {14},
  pages   = {9153},
  year    = {2024},
  doi     = {10.1038/s41598-024-59652-w}
}

@inproceedings{zhao2021plug,
  title     = {Plug-and-Play Domain Adaptation for Cross-Subject {EEG}-Based Emotion Recognition},
  author    = {Zhao, Li-Ming and Yan, Rui and Lu, Bao-Liang},
  booktitle = {Proceedings of the AAAI Conference on Artificial Intelligence},
  volume    = {35},
  pages     = {1521--1529},
  year      = {2021},
  doi       = {10.1609/aaai.v35i1.16169}
}

@article{jeon2023mutual,
  title   = {Mutual Information-Driven Subject-Invariant and Class-Relevant Deep Representation Learning in {BCI}},
  author  = {Jeon, Eunjin and Ko, Wonjun and Yoon, Jee Seok and Suk, Heung-Il},
  journal = {IEEE Transactions on Neural Networks and Learning Systems},
  volume  = {34},
  number  = {2},
  pages   = {739--749},
  year    = {2023},
  doi     = {10.1109/TNNLS.2021.3100583}
}

@article{charles2019mental,
  title={Measuring mental workload using physiological measures: A systematic review},
  author={Charles, Rebecca L and Nixon, Jim},
  journal={Applied ergonomics},
  volume={74},
  pages={221--232},
  year={2019},
  publisher={Elsevier}
}

@article{bastos2016connectivity,
  title={A Tutorial Review of Functional Connectivity Analysis Methods and Their Interpretational Pitfalls},
  author={Bastos, Andr{\'e} M. and Schoffelen, Jan-Mathijs},
  journal={Frontiers in Systems Neuroscience},
  volume={9},
  pages={175},
  year={2016},
  doi={10.3389/fnsys.2015.00175}
}

@article{dimitriadis2015workload,
  title={Cognitive workload assessment based on the tensorial treatment of {EEG} estimates of cross-frequency phase interactions},
  author={Dimitriadis, Stavros I. and Sun, Yu and Kwok, Kenneth and Laskaris, Nikolaos A. and Thakor, Nitish and Bezerianos, Anastasios},
  journal={Annals of Biomedical Engineering},
  volume={43},
  number={4},
  pages={977--989},
  year={2015},
  doi={10.1007/s10439-014-1143-0}
}

@inproceedings{zhang2017graphworkload,
  title={Graph theoretical analysis of {EEG} functional network during multi-workload flight simulation experiment in virtual reality environment},
  author={Zhang, Shengqian and Zhang, Yuan and Sun, Yu and Thakor, Nitish and Bezerianos, Anastasios},
  booktitle={2017 39th Annual International Conference of the IEEE Engineering in Medicine and Biology Society},
  pages={3957--3960},
  year={2017},
  doi={10.1109/EMBC.2017.8037722}
}

@inproceedings{miao2022gsat,
  title={Interpretable and Generalizable Graph Learning via Stochastic Attention Mechanism},
  author={Miao, Siqi and Liu, Mia and Li, Pan},
  booktitle={Proceedings of the 39th International Conference on Machine Learning},
  series={Proceedings of Machine Learning Research},
  volume={162},
  pages={15524--15543},
  publisher={PMLR},
  year={2022}
}

@inproceedings{maddison2017concrete,
  title={The Concrete Distribution: A Continuous Relaxation of Discrete Random Variables},
  author={Maddison, Chris J. and Mnih, Andriy and Teh, Yee Whye},
  booktitle={International Conference on Learning Representations},
  year={2017}
}

@book{chung1997spectral,
  title={Spectral Graph Theory},
  author={Chung, Fan R. K.},
  series={CBMS Regional Conference Series in Mathematics},
  volume={92},
  publisher={American Mathematical Society},
  address={Providence, Rhode Island},
  year={1997}
}

@article{Wang_ARFN_2024,
  title     = {{ARFN}: An Attention-Based Recurrent Fuzzy Network for
               {EEG} Mental Workload Assessment},
  author    = {Wang, Zhengyi and Ouyang, Yu and Zeng, Hong},
  journal   = {IEEE Transactions on Instrumentation and Measurement},
  volume    = {73},
  pages     = {1--14},
  year      = {2024},
  doi       = {10.1109/TIM.2024.3369143}
}

@inproceedings{Weng_StateMamba_2026,
  title     = {State {Mamba}: Spatiotemporal {EEG} State-Space Model
               with Dynamic Brain Alignment for Cross-Subject
               Representation},
  author    = {Weng, Weining and Gu, Yang and Ma, Yuan and
               Liu, Yuchen and Zhang, Yingwei and Chen, Yiqiang},
  booktitle = {Proceedings of the AAAI Conference on Artificial
               Intelligence},
  pages     = {17850--17858},
  year      = {2026},
  doi       = {10.1609/aaai.v40i21.38843}
}

\end{document}